\documentclass{PoS}

\usepackage{amssymb}
\usepackage{amsthm}
\usepackage{amsmath}
\usepackage{graphicx}
\usepackage{subcaption}
\newcommand{\trento}{T\raisebox{-0.5ex}{R}ENTo}

\newcommand{\be}{\begin{eqnarray}}
\newcommand{\ee}{\end{eqnarray}}

\newcommand{\ml}{\mathcal}
\newcommand{\bs}{\boldsymbol}

\title{Quarkonium production in heavy ion collisions: coupled Boltzmann transport equations}

\ShortTitle{Quarkonium production}

\author{\speaker{Xiaojun Yao}%\thanks{}
,  {Weiyao Ke}, {Yingru Xu}, {Steffen Bass} and {Berndt M\"uller}\\
        Department of Physics, Duke University, NC 27708 U.S.\\
        E-mail: \email{xiaojun.yao@duke.edu}}

\abstract{By coupling the Boltzmann transport equations of both quarkonium and open heavy quarks, we investigate their dynamical evolution inside the quark-gluon plasma and study quarkonium production in heavy ion collisions. The Boltzmann transport equation of quarkonium is derived from the open quantum system formalism and effective field theory of QCD by assuming quarkonium interacts weakly with the plasma. The dissociation and recombination terms in the Boltzmann equation are calculated in potential nonrelativistic QCD. It is shown that the combination of quarkonium dissociation, recombination, open heavy quark diffusion and energy loss can drive the system of quarkonium and open heavy quarks to detailed balance and kinetic thermalization. By solving the transport equations with initial momenta of quarkonia and heavy quarks sampled from \textsc{Pythia} and a hydrodynamic medium, we can calculate the nuclear modification factors of bottomonium and describe the data at both RHIC and LHC energies. The azimuthal angular anisotropy coefficient $v_2$ of $\Upsilon$(1S) in 5.02 TeV peripheral Pb-Pb collisions is also predicted.}

\FullConference{International Conference on Hard and Electromagnetic Probes of High-Energy Nuclear Collisions\\
		30 September - 5 October 2018\\
		Aix-Les-Bains, Savoie, France}

\begin{document}

\section{Introduction}
Since the early study of the static plasma screening effect on heavy quark bound states, quarkonium has been used as a probe of quark-gluon plasma (QGP) in heavy ion collisions \cite{Matsui:1986dk}. In addition to the static screening effect that suppresses the attraction between the heavy quark antiquark pair, there is also the dynamical screening effect, which is the dissociation of quarkonium due to collisions with medium constituents in the QGP. The dissociation rate of quarkonium has been widely investigated, by using different methods in both weak and strong coupling limits. The inverse process, recombination of unbound heavy quark antiquark pairs, can also occur inside QGP. The recombination has been modeled phenomenologically using the detailed balance at equilibrium. However, the recombination should depend on the distribution of heavy quark antiquark pairs that evolve with time inside the QGP and the distributions are not necessarily thermal in the early stage of the evolution.

We develop a rigorous treatment of recombination and put the dissociation and recombination in the same theoretical framework. To this end, we couple the Boltzmann transport equations of quarkonium with those of open heavy quarks. The transport equations of open heavy quarks provide the real-time distributions of open heavy quarks that are used when computing the recombination. Both the dissociation and recombination terms in the transport equation are calculated from potential nonrelativistic QCD (pNRQCD) \cite{Brambilla:1999xf}. In Section~\ref{sect:boltzmann}, the set of coupled Boltzmann equations is introduced and the calculation of dissociation and recombination terms is briefly explained. Then in Section~\ref{sect:solve}, methods used to simulate the transport equations are explained. Comparisons with experimentally measured nuclear modification factors $R_{AA}$ are shown in Section~\ref{sect:results}. Finally, a short conclusion is drawn in Section~\ref{sect:conclusion}.

\section{Coupled Boltzmann transport equations}
\label{sect:boltzmann}
The Boltzmann transport equation of quarkonium can be derived from first principles by assuming quarkonium interacts weakly with the plasma \cite{Yao:2018nmy}. The derivation uses the open quantum system formalism and effective field theory pNRQCD. When a Wigner transform is applied to the Lindblad equation, it leads to a Boltzmann transport equation under the Markovian approximation. 

The coupled Boltzmann equations of open heavy quarks $Q$, antiquarks $\bar{Q}$ and each quarkonium state with the quantum number $nls$ are given by \cite{Yao:2018zrg} 
\be
\label{eq:LBE}
(\frac{\partial}{\partial t} + \dot{{\bs x}}\cdot \nabla_{\bs x})f_Q({\bs x}, {\bs p}, t) &=& \ml{C}_Q  -  \ml{C}_Q^{+} +  \ml{C}_Q^{-}\\
(\frac{\partial}{\partial t} + \dot{{\bs x}}\cdot \nabla_{\bs x})f_{\bar{Q}}({\bs x}, {\bs p}, t) &=& \ml{C}_{\bar{Q}}  -  \ml{C}_{\bar{Q}}^{+} + \ml{C}_{\bar{Q}}^{-}\\
(\frac{\partial}{\partial t} + \dot{{\bs x}}\cdot \nabla_{\bs x})f_{nls}({\bs x}, {\bs p}, t) &=& \ml{C}_{nls}  +  \ml{C}_{nls}^{+}-\ml{C}_{nls}^{-} \,,
\ee
where $f_i({\bs x}, {\bs p}, t)$ denotes the phase space distribution function for $i=Q$, $\bar{Q}$ or each quarkonium state $nls$. Here $n=1$ is the ground state and higher $n$'s are excited states, $l$ is the angular momentum and $s$ is the spin. In our calculations, we will average over the polarizations of non-S wave quarkonium states so we omit the dependence on the quantum number $m$. 
The left-hand side of the equations describes the free streaming of particles while the right-hand side contains interactions between heavy particles and light quarks and gluons (abbreviated as $q$ and $g$) in the hot medium.

The collision term $\ml{C}_i$ for the interaction between the open heavy-flavor and the medium includes both elastic and inelastic contributions with full detailed balance \cite{Ke:2018tsh,Ke:2018jem}. For quarkonium, the elastic scattering $g+H \leftrightarrow g+H$ ($H$ indicates a quarkonium state) can contribute to $\ml{C}$. It has been calculated and shown to be negligible \cite{Yao:2018sgn}. In the language of effective field theory, the elastic scattering process between quarkonium and gluon occurs at the order $r^2$, where $r$ is the typical size of the quarkonium state. When the quarkonium is smaller in size than $1/T$, the elastic scattering is suppressed with respect to the dissociation and recombination, which happen at the order $r$.

The terms $\ml{C}_i^\pm$ describe the recombination and dissociation of quarkonium. Contributing processes include the gluon absorption/emission $g+H \leftrightarrow Q + \bar{Q}$ and inelastic scattering with medium constituents $g(q)+H \leftrightarrow g(q)+ Q + \bar{Q}$. The explicit expressions of $\ml{C}_i^\pm$ have been written out by computing the scattering amplitudes of relevant processes in pNRQCD \cite{Yao:2018sgn}. These scattering amplitudes have been shown to satisfy the Ward identities. The inelastic scattering is a $t$-channel process. It has also been shown that the $t$-channel process is infrared safe. The finite binding energy of quarkonium serves as a soft divergence regulator. The interference between the gluon absorption/emission and its thermal loop corrections cancels the collinear divergence in the $t$-channel process. See Ref.~\cite{Yao:2018sgn} for detailed discussions. In this work, we use Coulomb potential to calculate bound state wavefunctions.

\section{Solving transport equations}
\label{sect:solve}
We solve the coupled Boltzmann equations by Monte Carlo simulations. The medium background is provided by a boost-invariant viscous hydrodynamic simulation \cite{Shen:2014vra}. \trento\, provides the averaged initial entropy density for the hydrodynamic simulation in each centrality class \cite{Moreland:2014oya}. The medium properties used in the hydrodynamic simulation have been calibrated with experimental observables of light particles with small transverse momenta \cite{Bernhard:2016tnd}.

In the initialization, we sample a certain number of quarkonia and heavy quarks from the initial distribution functions. Initial momenta are sampled from the event generator \textsc{Pythia} \cite{Sjostrand:2014zea} with the nuclear parton distribution functions (PDF) EPS09 \cite{Eskola:2009uj}. The initial production suppression due to the cold nuclear matter effect is accounted by the nuclear PDF. 
Initial positions are sampled from the averaged binary collision density profile calculated in the \trento\, model.

At each time step of the simulation, we consider three types of processes: diffusion and radiation energy loss of open heavy quarks, dissociation of quarkonium and recombination of open heavy quark antiquark pairs. 
The probability for each process to happen within a time step is calculated from the respective reaction rate. If a process happened according to its reaction probability, the momenta of final-states are sampled from the differential reaction rate in a Monte Carlo way that conserves both energy and momentum. For dissociation, we replace the dissociating quarkonium with an unbound heavy quark antiquark pair. Both heavy quarks have the same position as the quarkonium before the dissociation. For recombination, we replace the pair of heavy quark and antiquark with a specific quarkonium state, whose position is given by the center-of-mass position of the pair before recombination. If the local QGP temperature is higher than the melting temperature of a specific quarkonium state, recombination into that quarkonium state is forbidden.

Before carrying out simulations for real collisions, we test the simulation inside a QGP box with periodic boundary conditions and a constant temperature throughout the box. It has been demonstrated that the interplay between the dissociation and recombination drives the system into detailed balance \cite{Yao:2017fuc}. Since the diffusion term of quarkonium is neglected in the Boltzmann equation, quarkonium reaches kinetic thermalization via recombinations of heavy quarks that are kinetically thermalized.

\section{Results}
\label{sect:results}
\begin{figure}[h]
    \centering
        \begin{subfigure}[t]{0.5\textwidth}
        \centering
        \includegraphics[height=1.8in]{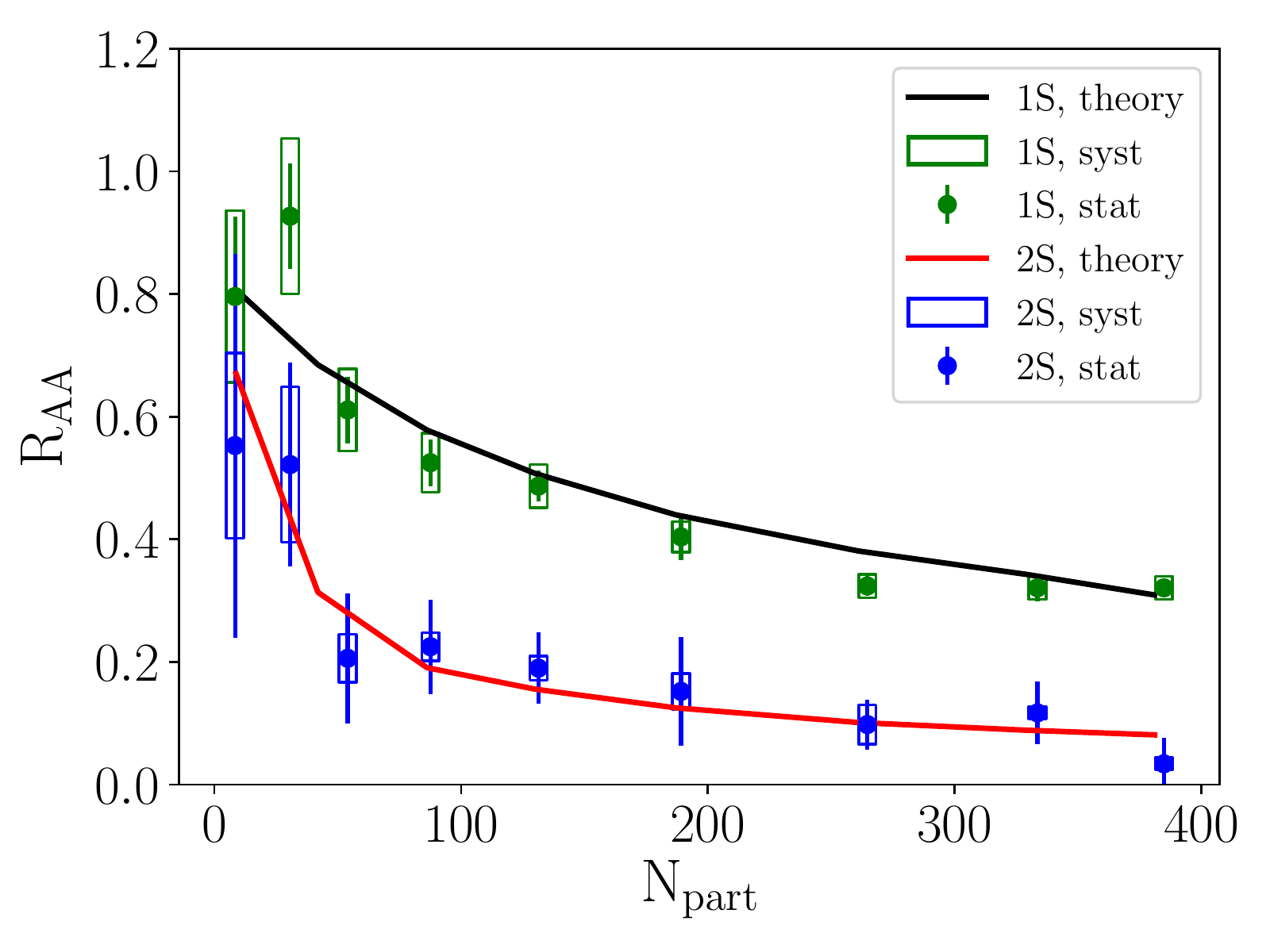}
        \caption{$R_{AA}$ as a function of centrality.}
    \end{subfigure}%
    ~
    \begin{subfigure}[t]{0.5\textwidth}
        \centering
        \includegraphics[height=1.8in]{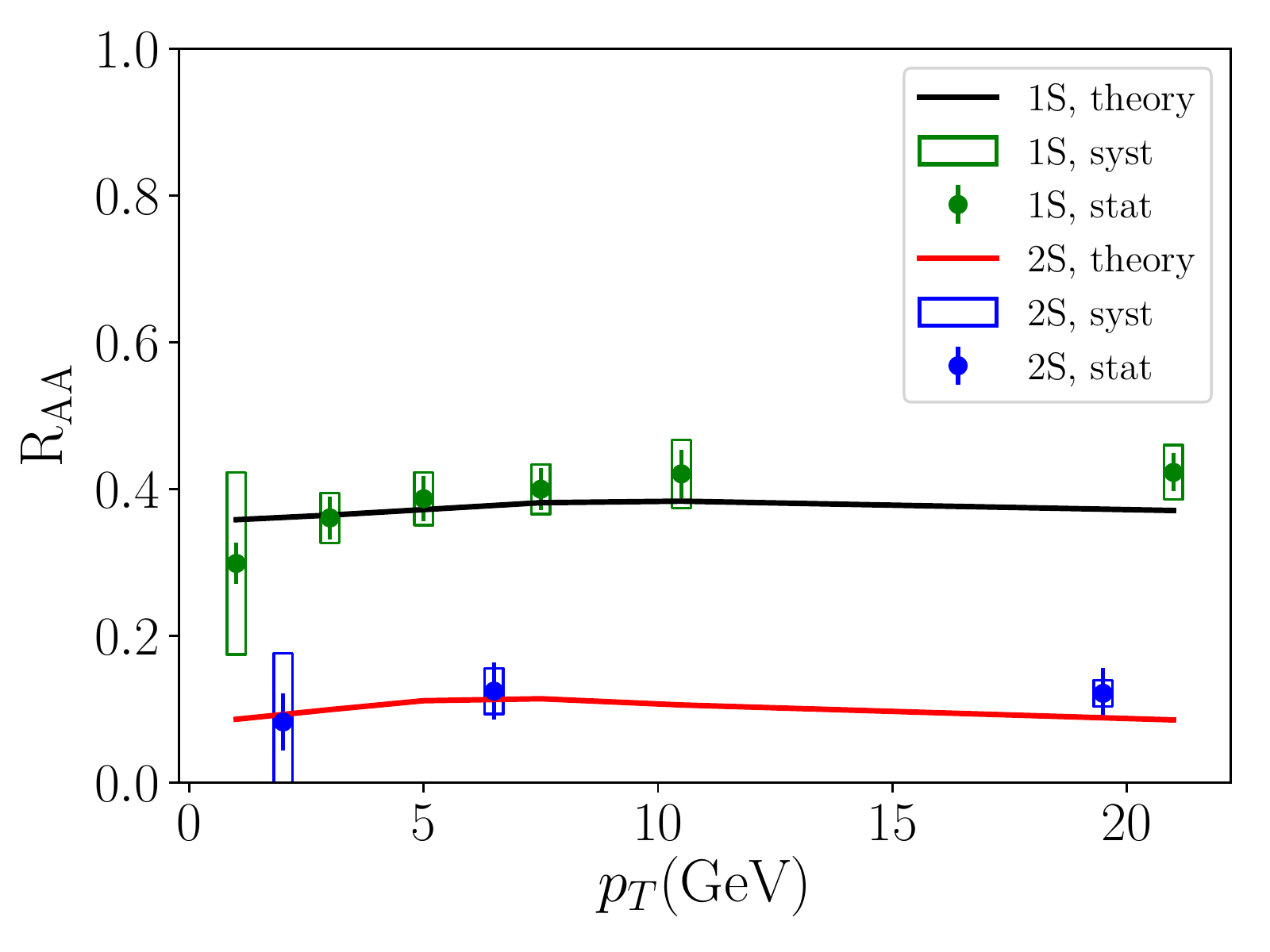}
        \caption{$R_{AA}$ as a function of transverse momentum.}
    \end{subfigure}%
    
    \begin{subfigure}[t]{0.5\textwidth}
        \centering
        \includegraphics[height=1.8in]{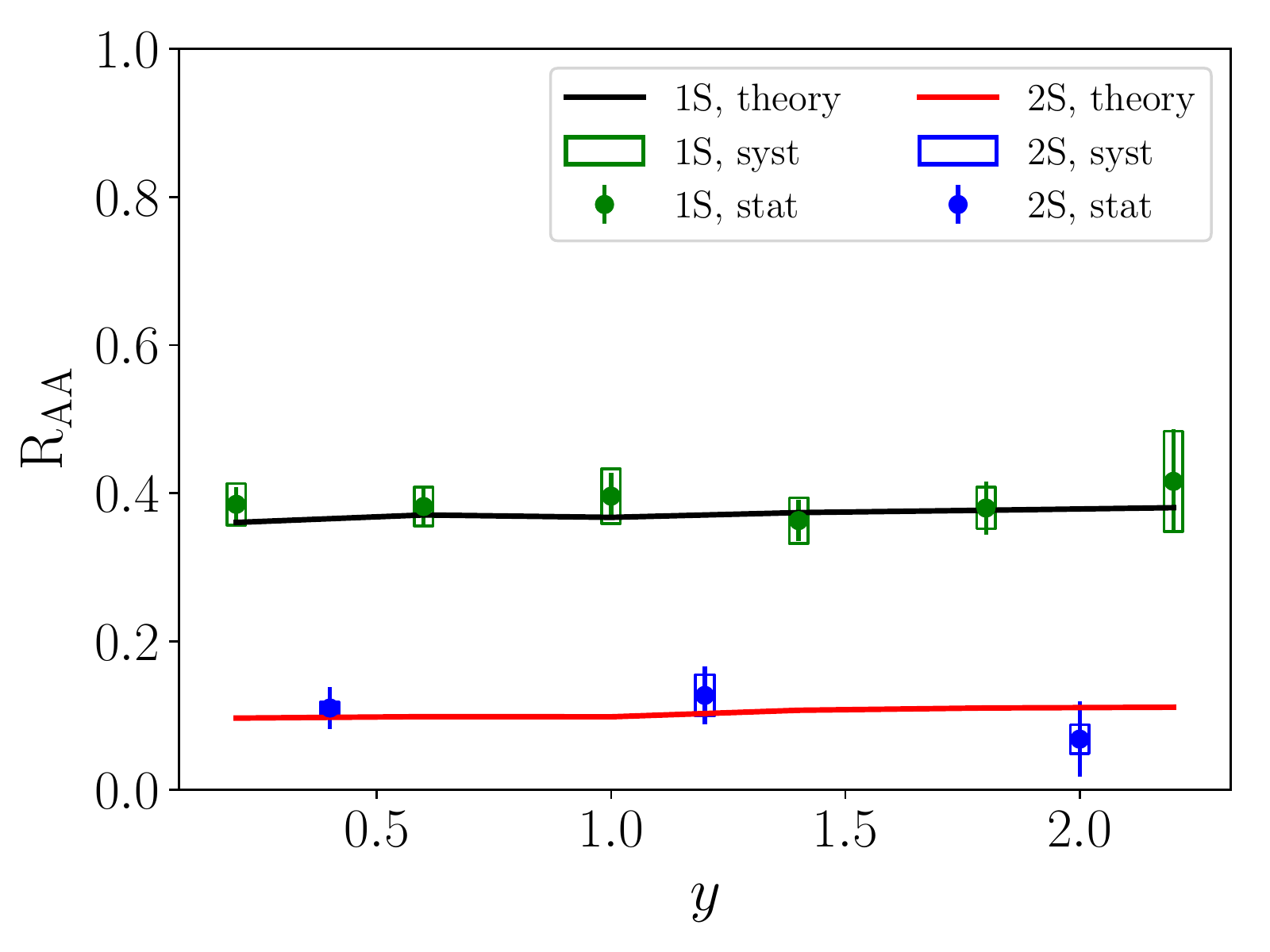}
        \caption{$R_{AA}$ as a function of rapidity.}
    \end{subfigure}%
    ~
    \begin{subfigure}[t]{0.5\textwidth}
        \centering
        \includegraphics[height=1.8in]{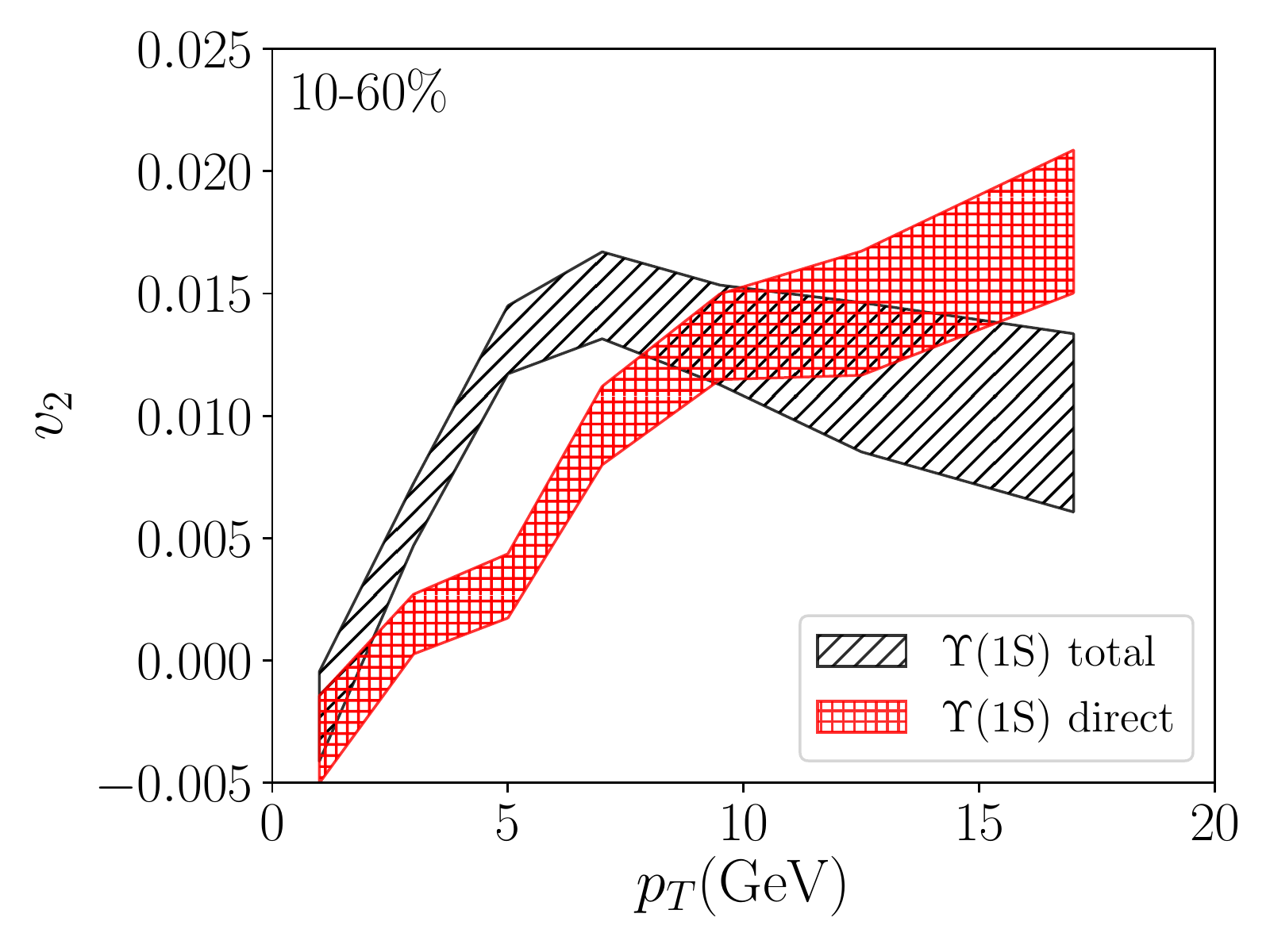}
        \caption{$v_2$ of $\Upsilon$(1S) in the centrality range $10\%-60\%$.}
    \end{subfigure}%
    \caption{Results of $R_{AA}$ and $v_2$ in $5.02$ TeV Pb-Pb collisions and CMS measurements of $R_{AA}$. Data are taken from Ref.~\cite{Sirunyan:2018nsz}}
    \label{fig:cms}
\end{figure}

Here we focus on the Upsilon production and include both 1S and 2S states in the calculation. Due to the larger bottom quark mass, results in the leading order expansion of $r$ are more reliable for bottomonium than charmonium. By choosing the coupling constant, the singlet potential, and the melting temperature of $\Upsilon$(2S) as $\alpha_s=0.3$, $V_s = -C_F\frac{0.42}{r}$, and $210$ MeV respectively, we can describe the Upsilon $R_{AA}$ in $200$ GeV Au-Au collisions measured by the STAR collaboration and that in $2.76$ TeV Pb-Pb collisions measured by the CMS collaboration. The results have been reported in Ref.~\cite{Yao:2018zrg}. Here we use the same set of parameters and do the calculation for $5.02$ TeV Pb-Pb collisions. The results are shown in Fig.~\ref{fig:cms}. Our results are consistent with the CMS measurements. Productions in central collisions are more suppressed than in peripheral collisions. Suppressions are nearly independent of the transverse momentum and rapidity. We also study the azimuthal angular anisotropy of $\Upsilon$(1S). We compare the total $v_2$ of $\Upsilon$(1S) with the $v_2$ of those $\Upsilon$(1S) that are produced directly in the initial hard scattering and survive the in-medium evolution. In addition to the direct production, recombination production also contributes to $v_2$. The $v_2$ of recombined $\Upsilon$(1S) is expected to be large at low transverse momentum and decrease at large transverse momentum because of the $v_2$ of recombining bottom quarks. As a result, the total $v_2$ decreases at large transverse momentum.

\section{Conclusions}
\label{sect:conclusion}

We use a set of coupled Boltzmann equations to describe the in-medium evolution of heavy quarks and quarkonia. By choosing appropriate parameter values, we can explain the experimental data of Upsilon $R_{AA}$ at both RHIC and LHC energies. An estimate of the $\Upsilon$(1S) $v_2$ in $5.02$ TeV Pb-Pb collisions is presented. The same framework can be modified to study the production of doubly charm baryon in heavy ion collisions \cite{Yao:2018zze}. In future, we plan to use potentials motivated from lattice calculations when computing quarkonium wavefunctions, include higher excited quarkonium states in the transport equations and use event-by-event hydrodynamic simulations to give a more complete understanding of quarkonium production in heavy ion collisions.

The work is supported by U.S. Department of Energy under Research Grant No. DE-FG02-05ER41367. X.Y. also acknowledges support from Brookhaven National Laboratory.


\begin{thebibliography}{99}
%\cite{Matsui:1986dk}
\bibitem{Matsui:1986dk} 
  T.~Matsui and H.~Satz,
  %``$J/\psi$ Suppression by Quark-Gluon Plasma Formation,''
  Phys.\ Lett.\ B {\bf 178}, 416 (1986).
  %doi:10.1016/0370-2693(86)91404-8
  %%CITATION = doi:10.1016/0370-2693(86)91404-8;%%
  %2755 citations counted in INSPIRE as of 25 Jun 2018

\bibitem{Brambilla:1999xf} 
  N.~Brambilla, A.~Pineda, J.~Soto and A.~Vairo,
  %``Potential NRQCD: An Effective theory for heavy quarkonium,''
  Nucl.\ Phys.\ B {\bf 566}, 275 (2000)
  %doi:10.1016/S0550-3213(99)00693-8
  [hep-ph/9907240].
  %%CITATION = doi:10.1016/S0550-3213(99)00693-8;%%
  %481 citations counted in INSPIRE as of 10 Aug 2017
    
%\cite{Yao:2018nmy}
\bibitem{Yao:2018nmy} 
  X.~Yao and T.~Mehen,
  %``Quarkonium in-Medium Transport Equation Derived from First Principles,''
  arXiv:1811.07027 [hep-ph].
  %%CITATION = ARXIV:1811.07027;%%

%\cite{Yao:2018zrg}
\bibitem{Yao:2018zrg} 
  X.~Yao, W.~Ke, Y.~Xu, S.~Bass and B.~Müller,
  %``Quarkonium production in heavy ion collisions: coupled Boltzmann transport equations,''
  arXiv:1807.06199 [nucl-th].
  %%CITATION = ARXIV:1807.06199;%%
  %1 citations counted in INSPIRE as of 27 Nov 2018

%\cite{Ke:2018tsh}
\bibitem{Ke:2018tsh} 
  W.~Ke, Y.~Xu and S.~A.~Bass,
  %``A linearized Boltzmann--Langevin model for heavy quark transport in hot and dense QCD matter,''
  arXiv:1806.08848 [nucl-th].
  %%CITATION = ARXIV:1806.08848;%%
  %4 citations counted in INSPIRE as of 27 Nov 2018

%\cite{Ke:2018jem}
\bibitem{Ke:2018jem} 
  W.~Ke, Y.~Xu and S.~A.~Bass,
  %``Modeling of quantum-coherence effects in parton radiative energy loss,''
  arXiv:1810.08177 [nucl-th].
  %%CITATION = ARXIV:1810.08177;%%
      
%\cite{Yao:2018sgn}
\bibitem{Yao:2018sgn} 
  X.~Yao and B.~Müller,
  %``Quarkonium inside Quark-gluon Plasma: Diffusion, Dissociation, Recombination and Energy Loss,''
  arXiv:1811.09644 [hep-ph].
  %%CITATION = ARXIV:1811.09644;%%

%\cite{Shen:2014vra}
\bibitem{Shen:2014vra} 
  C.~Shen, Z.~Qiu, H.~Song, J.~Bernhard, S.~Bass and U.~Heinz,
  %``The iEBE-VISHNU code package for relativistic heavy-ion collisions,''
  Comput.\ Phys.\ Commun.\  {\bf 199}, 61 (2016).
  %doi:10.1016/j.cpc.2015.08.039
  %[arXiv:1409.8164 [nucl-th]].
  %%CITATION = doi:10.1016/j.cpc.2015.08.039;%%
  %131 citations counted in INSPIRE as of 27 Jun 2018

%\cite{Moreland:2014oya}
\bibitem{Moreland:2014oya} 
  J.~S.~Moreland, J.~E.~Bernhard and S.~A.~Bass,
  %``Alternative ansatz to wounded nucleon and binary collision scaling in high-energy nuclear collisions,''
  Phys.\ Rev.\ C {\bf 92}, no. 1, 011901 (2015)
  %doi:10.1103/PhysRevC.92.011901
  [arXiv:1412.4708 [nucl-th]].
  %%CITATION = doi:10.1103/PhysRevC.92.011901;%%
  %78 citations counted in INSPIRE as of 27 Nov 2018

%\cite{Bernhard:2016tnd}
\bibitem{Bernhard:2016tnd} 
  J.~E.~Bernhard, J.~S.~Moreland, S.~A.~Bass, J.~Liu and U.~Heinz,
  %``Applying Bayesian parameter estimation to relativistic heavy-ion collisions: simultaneous characterization of the initial state and quark-gluon plasma medium,''
  Phys.\ Rev.\ C {\bf 94}, no. 2, 024907 (2016).
  %doi:10.1103/PhysRevC.94.024907
  %[arXiv:1605.03954 [nucl-th]].
  %%CITATION = doi:10.1103/PhysRevC.94.024907;%%
  %111 citations counted in INSPIRE as of 27 Jun 2018 

%\cite{Sjostrand:2014zea}
\bibitem{Sjostrand:2014zea} 
  T.~Sj\"ostrand {\it et al.},
  %``An Introduction to PYTHIA 8.2,''
  Comput.\ Phys.\ Commun.\  {\bf 191}, 159 (2015)
  %doi:10.1016/j.cpc.2015.01.024
  [arXiv:1410.3012 [hep-ph]].
  %%CITATION = doi:10.1016/j.cpc.2015.01.024;%%
  %1021 citations counted in INSPIRE as of 27 Jun 2018

%\cite{Eskola:2009uj}
\bibitem{Eskola:2009uj} 
  K.~J.~Eskola, H.~Paukkunen and C.~A.~Salgado,
  %``EPS09: A New Generation of NLO and LO Nuclear Parton Distribution Functions,''
  JHEP {\bf 0904}, 065 (2009)
  %doi:10.1088/1126-6708/2009/04/065
  [arXiv:0902.4154 [hep-ph]].
  %%CITATION = doi:10.1088/1126-6708/2009/04/065;%%
  %845 citations counted in INSPIRE as of 27 Jun 2018

%\cite{Yao:2017fuc}
\bibitem{Yao:2017fuc} 
  X.~Yao and B.~Müller,
  %``Approach to equilibrium of quarkonium in quark-gluon plasma,''
  Phys.\ Rev.\ C {\bf 97}, no. 1, 014908 (2018)
  Erratum: [Phys.\ Rev.\ C {\bf 97}, no. 4, 049903 (2018)]
  %doi:10.1103/PhysRevC.97.049903, 10.1103/PhysRevC.97.014908
  [arXiv:1709.03529 [hep-ph]].
  %%CITATION = doi:10.1103/PhysRevC.97.049903, 10.1103/PhysRevC.97.014908;%%
  %12 citations counted in INSPIRE as of 27 Nov 2018
        
%\cite{Sirunyan:2018nsz}
\bibitem{Sirunyan:2018nsz} 
  A.~M.~Sirunyan {\it et al.} [CMS Collaboration],
  %``Measurement of nuclear modification factors of $\Upsilon$(1S), $\Upsilon$(2S), and $\Upsilon$(3S) mesons in PbPb collisions at $\sqrt{s_{_\mathrm{NN}}} =$ 5.02 TeV,''
  arXiv:1805.09215 [hep-ex].
  %%CITATION = ARXIV:1805.09215;%%
  %9 citations counted in INSPIRE as of 27 Nov 2018

%\cite{Yao:2018zze}
\bibitem{Yao:2018zze} 
  X.~Yao and B.~Müller,
  %``Doubly charmed baryon production in heavy ion collisions,''
  Phys.\ Rev.\ D {\bf 97}, no. 7, 074003 (2018)
  %doi:10.1103/PhysRevD.97.074003
  [arXiv:1801.02652 [hep-ph]].
  %%CITATION = doi:10.1103/PhysRevD.97.074003;%%
  %7 citations counted in INSPIRE as of 27 Nov 2018
     
\end{thebibliography}
\end{document}